\documentclass[
 reprint,
 amsmath,amssymb,
 aps,
]{revtex4-2}

\usepackage{placeins} 
\usepackage{url}
\usepackage[hyperindex,breaklinks]{hyperref}
\hypersetup{linkcolor=blue, citecolor=red, colorlinks=true}

\usepackage{capt-of}
\usepackage{graphicx}
\usepackage{dcolumn}
\usepackage{bm}

\begin{document}

\title{
Parrondo's effect in continuous-time quantum walks 
}

\author{J. J. Ximenes$^{1}$}
\thanks{jeffersonximenes@ufu.br}

\author{M. A. Pires$^{2}$}
\thanks{marcelo.pires@delmiro.ufal.br}

\author{J. M. Villas-Bôas$^{1}$}
\thanks{boas@ufu.br}

\affiliation{
$^{1}$Instituto de Física, Universidade Federal de Uberlândia, 38400-902 Uberlândia-MG, Brazil
\\ 
$^{2}$Universidade Federal de Alagoas, 57480-000, Delmiro Gouveia - AL, Brazil
}

\begin{abstract}
We present the first manifestation of a Parrondo's effect in a 
 continuous-time quantum walk (CTQW). In our protocol we consider a  CTQW in the presence of time-dependent transition defect. Our results show that 
the alternation between defects, that individually are detrimental to the wavepacket spreading, can paradoxically enhance overall wavepacket propagation. 
 Our findings pave the way for the exploration of unconventional mechanisms that can potentially harness the adverse effects of defects to enhance quantum transport. 
\end{abstract}


\maketitle


\section{Introduction}\label{sec:intro}

The standard quantum walk (QW) was originally introduced in 1993~\cite{aharonov1993quantum} in a discrete-time formulation. Subsequently, in 1998, the continuous-time variant of this model was developed~\cite{farhi1998firstCTQW}. The QW is a lattice-based model exhibiting several interesting properties~\cite{kempe2003quantum} when compared with the classical random walk (CRW). 
Comparative analysis between the QW with the classical random walk (CRW) reveals compelling distinctions. 
Firstly, the spreading of the QW is quadratically faster than the CRW.
Secondly, the QW exhibits a non-Gaussian bimodal probability distribution, in stark contrast to the Gaussian distribution observed in the CRW between both models.
The contrast between  QWs and CRWs is further accentuated when both models 
spread on graphs~\cite{childs2002example}.
Recently, the Ref.~\cite{pires2020quantum} presented a new difference between the CRW and the QW: the introduction of short-range aperiodic jumps amplifies the spreading of CRWs, while, counterintuitively, it induces an inhibition of wavepacket spreading of QWs.

The wide versatility of QWs leads this  model to serve as a computational  platform for investigating a broad diversity of phenomena.
For instance, with QWs it is possible to explore 
topological phases~\cite{kitagawa2010exploring,wu2019topological}, solitonlike propagation~\cite{mendoncca2020emergent},  rogue waves~\cite{buarque2022rogue}, 
trojan's effects~\cite{ghizoni2019trojan},
the Ramsauer effect~\cite{lam2015ramsauer}, $q$-Gaussian distributions~\cite{shikano2014discrete}, Anderson localization~\cite{derevyanko2018anderson,ghosh2014simulating}, 
hyperballistic regimes~\cite{di2018elephant}, and multiple transitions between  diffusive, superdiffusive, ballistic, and hyperballistic behavior~\cite{pires2019multiple,naves2022enhancing,naves2023quantum}. 
QWs may also lead to intriguing phenomena~\cite{kendon2003decoherence} and  non-monotonic effects with decoherence~\cite{oliveira2006decoherence}. 
The substantial interest in QWs also stems from their numerous algorithmic applications~\cite{ambainis2003quantum,portugal2013quantum,venegas2008quantum,venegas2012quantum,kadian2021quantum} and diverse experimental implementations~\cite{wang2013physical}.
The extensive range of possibilities  offered by QWs underscores the significance of investigating novel protocols for such model.

In this work, we undertake a theoretical examination of a QW  on an infinite line with alternating defects. 
The subsequent sections of this manuscript are structured as follows:
in Sec.~\ref{sec:related} we review articles that are related to our work; 
in Sec.~\ref{sec:model} we introduce our quantum protocol; 
in Sec.~\ref{sec:results} we disclose our results and we discuss our findings considering several measures
and in Sec.~\ref{sec:remaks}  we present final considerations and further perspectives on our work.

\section{Related works}\label{sec:related}

We now direct our focus toward more specific works directly relevant to our investigation. The first subsection is devoted to an in-depth exploration of continuous-time quantum walks (CTQWs) and discrete-time quantum walks (DTQWs) with defects. In the second subsection, we review works that demonstrate the manifestation of the Parrondo's effect (PE) within the context of DTQWs, as there is no realization of such phenomenon in CTQWs, so far. 

\subsection{QWs with Defects}

The exploration of defects in QWs has evolved systematically, offering valuable insights into how they shape the wavepacket dynamics. The work of Childs et al.~\cite{childs2002example} offered early insights into how a single defect can disrupt wavepacket propagation of CTQWs on graphs, laying the groundwork for further investigations.
Zhang et al.~\cite{zhang2014one} observed the localization effect in 1D-CTQWs with single-point phase defects, contributing to the understanding of defect-induced phenomena. Keating et al.~\cite{keating2007localization} explored CTQWs with Cauchy-distributed defects, shedding light on the localization effects associated with specific defect distributions. Agliari et al.~\cite{agliari2010quantum} investigated CTQWs with traps placed in fractal structures, revealing the behavior of wavepackets in non-trivial environments. Izaac et al.~\cite{izaac2013continuous} conducted a study on 1D-CTQWs in the presence of multiple defects. Their results show the presence of resonance behavior. 
Benedetti et al.~\cite{benedetti2019continuous} addressed the utilization of CTQWs as quantum probes for characterizing defects and perturbations within network structures.
Li et al.~\cite{li2013position} conducted a comprehensive study on CTQWs with potential defects, as well as DTQWs with phase defects, considering single and double position defects on a one-dimensional lattice. 
Later, Li and Wang~\cite{li2015analytical}   analytically investigated a model that is equivalent to a scattering transmission of 1D-CTQW with defects. More recently, Teles and Amorim~\cite{da2021localization} studied how  defects in DTQWs affect the return probability of a quantum particle. 
Kiumi and Saito~\cite{kiumi2021eigenvalues} have provided analytical results for two-phase DTQWs with one defect.
All these studies have enriched our comprehension of defect-induced effects in various QW models.

\subsection{Parrondo's effect in QWs}

The PE is traditionally formulated in terms of a combination of losing games that can produce a winning game~\cite{Parrondo1996,HarmerAbbott1999}. Over the years such a phenomenon 
has been observed in such a wide variety of 
fields~\cite{abbott2010asymmetry,cheong2019paradoxical,laiparrondo2020Review} that today the PE can be defined more generally as the emergence of favorable outcomes from 
combinations of unfavorable scenarios.

The earliest attempts to establish Parrondian QWs  can be traced to  Refs.~\cite{MeyerBlumer2002JSP,MeyerBlumer2002FNL,meyer2003noisy}.  These works were successful in the short-time, however they fail to obtain a stable   PE  in the long-time.
Similar efforts to introduce a Parrondo's paradox within QWs did not succeed in the asymptotic limit~\cite{Flitney2012,li2013quantum}.
Nowadays, there are several protocols for obtaining 
the Parrondo's effect or 
Parrondo-like effects in 
DTQWs~\cite{flitney2004quantum,gawron2005quantum,Kosik2007,chandrashekar2011parrondo,banerjee2013enhancement,Rajendran2018EPL,rajendran2018implementing,MachidaGrunbaum2018,walczak2023noise,walczak2022parrondo,walczak2021parrondo,trautmann2022parrondo,lai2020parrondoCointoss,lai2020parrondo4sided,mielke2023quantum,pires2020parrondo,panda2022generating,jan2023territories,janexperimental}.
Notably, in Refs.~\cite{pires2020parrondo,panda2022generating,jan2023territories} the authors have shown 
scenarios in which the PE in QWs are associated with an enhancement in  the corresponding coin-space entanglement between the internal (spin) and external (position) degrees of freedom.
In Ref.~\cite{janexperimental} it is presented the first experimental verification of a quantum Parrondo walk within a quantum optics setup.

\section{Model} \label{sec:model}

In this section, we describe step by step our quantum protocol.  

\subsection{CTQW}

 The Hamiltonian governing the dynamics of a CTQW for a single quantum particle, quantum walker, moving only between nearest neighbor sites within a uniform one-dimensional lattice can be expressed as
\begin{equation}\label{H0}
    H_0 = \epsilon \sum_{j} |j \rangle \langle j| - \gamma \sum_{j} \left( |j + 1 \rangle \langle j| + |j - 1 \rangle \langle j| \right) ,
\end{equation}
where $\epsilon$ represents the constant potential energy, while $\gamma$ denotes the transition rate. Thus, given an initial state $ | \Psi(t=0) \rangle $, the evolution of the system  can be described by the equation
\begin{equation}\label{Psit}
    i\frac{\partial}{\partial t} | \Psi(t) \rangle = H_0 | \Psi(t) \rangle  ,
\end{equation}
in which we set $\hbar=1$. The evolution of the system is assessed through the calculation of the probability distribution 
\begin{equation}\label{Pj}
P_j (t) = | \langle j | \psi (t) \rangle |^2 .
\end{equation}
To quantify the rate of spreading during this propagation, we analyze the standard deviation
\begin{equation}\label{std}
\sigma = \sqrt{\overline{j^2}- {\overline{j}}^2}  ,   
\end{equation}
where $\overline{j^n} = \langle \psi (t) | j^n | \psi (t) \rangle $.

\subsection{Transition defects} 

\begin{figure}[b]
    \centering
    \includegraphics[width=0.49\textwidth]{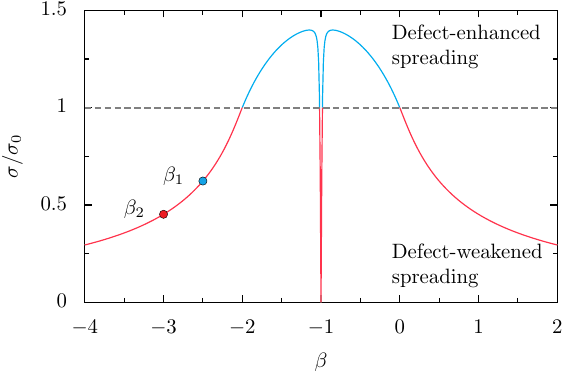}
    \caption{Relative standard deviation  $\sigma/\sigma_0$ as a function of $\beta$ for $\gamma t = 2000$. We use the reference $\sigma_0$ for the defect-free case. The dashed line separates the zone in which defects enhances spreading (above) from the region in which defects reduces spreading (below).}
    \label{fig:std_norm_vs_beta}
\end{figure}

The incorporation of transition defects within the lattice structure can be achieved by altering the transition rates between lattice sites. Following the approach of Li and Wang~\cite{li2015single}, in the context of a particle residing at site $j = d$ with nearest neighbors, the introduction of a transition defect is accomplished by the inclusion of the term:
\begin{equation}\label{Hb}
H_d = -  \left( |d \rangle \langle d+1| +|d + 1 \rangle \langle d| + |d - 1 \rangle \langle d|+ |d \rangle \langle d-1| \right) .
\end{equation}
This term is characterized by an associated transition rate denoted as $\beta$. The resultant modification of the Hamiltonian can be expressed as:
\begin{equation}\label{H}
    H_0 + \beta H_d  .
\end{equation}
In this study, we consider $\epsilon = 0$. 
Z. J. Li and J. B. Wang~\cite{li2015single} observed that for a specific value of $\beta = -0.5 \gamma$, the evolution of  $\sigma(t)$ surpasses that of the defect-free scenario. We first extend their results and we show in Fig.~\ref{fig:std_norm_vs_beta} that there is a nonmonotonic transition 
from the regime with  defect-weakened spreading  
to the regime with  defect-enhanced spreading.
The propagation of the particle is canceled for $ \beta = -\gamma $ because the transition rate becomes null $ \beta+\gamma = 0 $ in Eq.~\ref{H}; so that, from this point, symmetry is observed.
\subsection{Protocol for alternation}

Following the conventional mechanism of Parrondo's game, we implement an alternation between two scenarios with unfavorable outcomes, that in our model are characterized by weak spreading. To achieve this goal, a time-periodic function with a period of $T$, denoted as $f(t+T) = f(t)$, is defined such that, for $0 \leq t \leq T$,
\begin{equation}
    f(t) = \begin{cases}
		\beta_2, & \text{if $ t \leq \frac{T}{2} $ }\\
		\beta_1, & \text{if $ t > \frac{T}{2} $ }
           \end{cases}
\end{equation}
where the corresponding frequency is $ w=2 \pi/T $. 
By selecting two distinct transition rates, denoted as $ \beta_1 $ and $ \beta_2 $, the Hamiltonian governing this mechanism can be defined as:
\begin{equation}\label{Hp}
H = H_0 + f(t)H_d  .
\end{equation}
Consequently, the Hamiltonian takes on the form of Equation (\ref{H}) during intervals of time $T/2$. Within each of these intervals, the transition rate different, allowing for a switching between two distinct  modes.

\begin{figure}[t]
    \centering
    \includegraphics[width=0.49\textwidth]{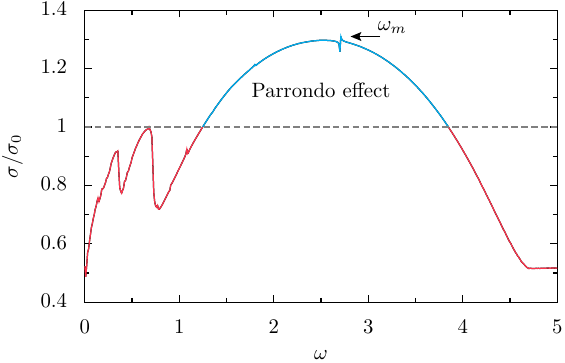}
\caption{Relative standard deviation $\sigma/\sigma_0$ as a function of $\beta$ for $\gamma t= 2000$.  The dashed line separates the zone in which defects enhances spreading (above) from the region in which defects reduces spreading (below).}
    \label{fig:sigmavsw}
\end{figure}

\begin{figure}[t]
    \centering
    \includegraphics[width=0.49\textwidth]{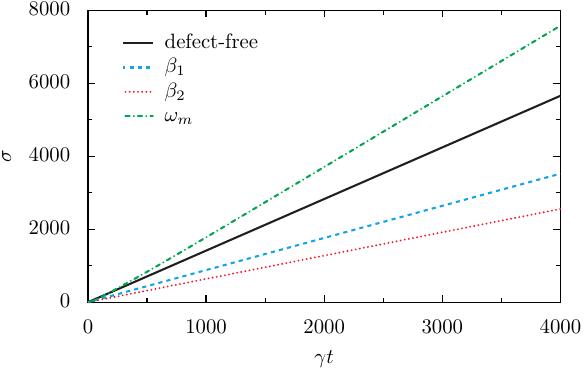}
\caption{Time evolution of the relative standard deviation $\sigma/\sigma_0$ for our Parrondian QW (with $\omega_m=2.71$), the defect-free QW as well as the 
    QW with defects (with $\beta_1 = -2.5 \gamma $ and  $\beta_2 = -3 \gamma $). The fastest spreading occurs for the Parrodian QW.}
    \label{fig:Standarvst}
\end{figure}

\begin{figure}[h]
    \centering
    \includegraphics[width=0.49\textwidth]{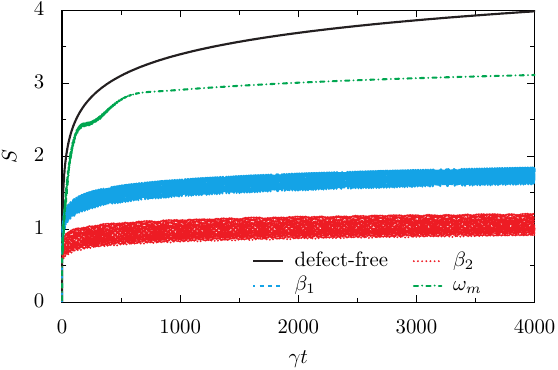}
\caption{Time evolution of the Shannon entropy $S$ for the same models and parameters presented in the Fig.~\ref{fig:Standarvst}. The defect-free QW exhibits the highest values for Shannon entropy.}
    \label{fig:Shannonvst}
\end{figure}

\begin{figure}[h]
    \centering
    \includegraphics[width=0.49\textwidth]{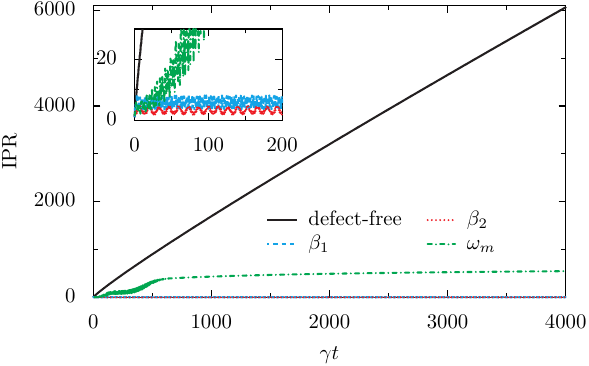}
\caption{Time evolution of the $IPR$ for the same models and parameters presented in the Fig.~\ref{fig:Standarvst}. The  $IPR$ of the defect-free QW presents the highest values.}
    \label{fig:IPRvst}
\end{figure}

\begin{figure*}[t]
    \centering
    \includegraphics[width=0.99\textwidth]{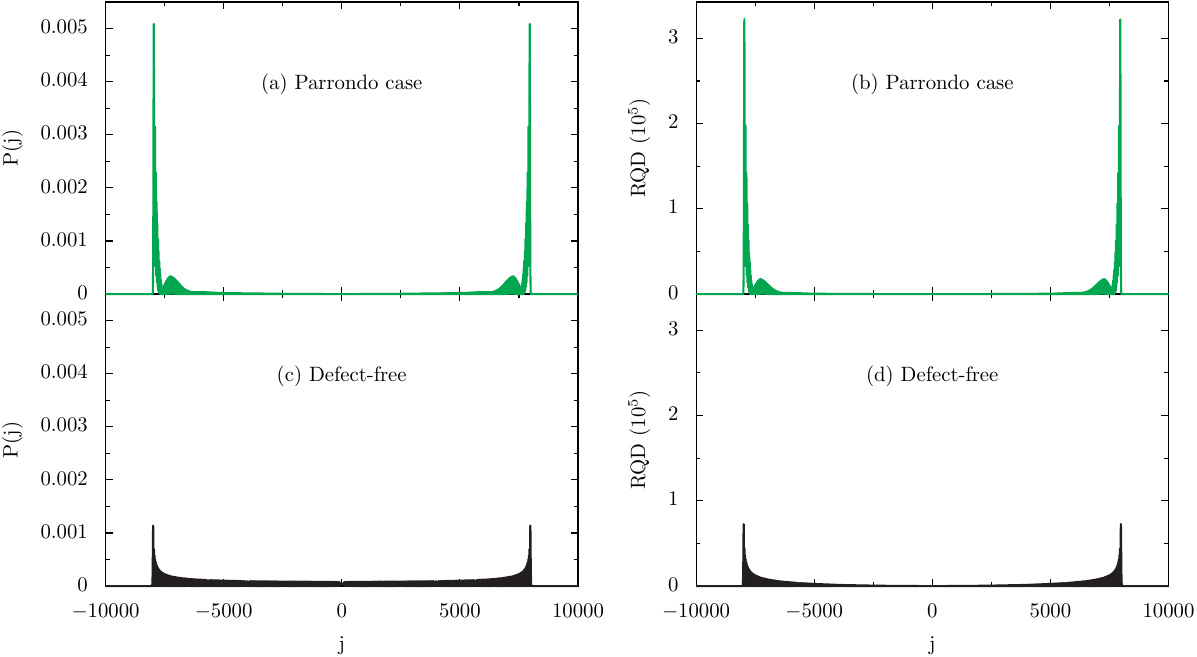}
\caption{Comparison of probability distribution and Relative Quadratic Deviation between case defect-free QW and Parrondian QW for $ \gamma t= 4000 $.}
    \label{fig:prqd}
\end{figure*}

\section{Results and discussion} \label{sec:results}

In this section, we present the results and discussion of our Parrondian QW. Our analysis involves a comprehensive comparison with the QW in which defects typically have a detrimental effect on wavepacket spreading (associated with $\beta_1$ and $\beta_2$ as illustrated in Fig.~\ref{fig:std_norm_vs_beta}). Additionally, we compare our results with the standard defect-free QW.

Following the model given by the Eq.~\ref{Hp}, we alternate the dynamics with $\beta_1$ and $\beta_2$ for a range of frequencies depicted in Fig.~\ref{fig:sigmavsw}. The measures $\sigma$ and $\sigma_0$ quantify the wavepacket spreading for scenarios with and without defects, respectively.
The dynamics characterized by  $\beta_1$ and $ \beta_2$ are selected, 
both exhibiting $\sigma/\sigma_0<1$ (reduced spreading).
The results show the emergence of an interesting  zone in which the Parrondo's paradox is detected. Specifically, it is possible to observe that the switching between two regimes with 
$\sigma/\sigma_0<1$ (weakened spreading) can lead to the appearance of a regime 
with $\sigma/\sigma_0>1$ (enhanced spreading).

In Fig.~\ref{fig:Standarvst}, we observe that the time evolution for the standard deviation of the Parrondo case
(with frequency $w_m$ as illustrated in Fig.~\ref{fig:sigmavsw}) surpasses that of the model with 
weak spreading induced by defects (for $\beta_1$ and $\beta_2$). Notably, 
the Parrondian QW also overcomes the standard deviation for the setting without defects.
Thus, our protocol provides a QW with alternating defects that can be tuned to exhibit an enhanced spreading rate compared to the usual QW model. We also have checked that
our model is still ballistic. That is, in contrast to previous investigations~\cite{di2018elephant,pires2019multiple,naves2022enhancing,naves2023quantum}, our accelerated QW model does not present a hyper-ballistic scaling.

To gain insights into the spatial distribution of wavepackets across the $N$ sites at a given instant of time, 
we calculate two distributional measures. First, we compute the Shannon entropy given by
\begin{equation}\label{Shannon}
    S = -\sum_j P_j \ln P_j  . 
\end{equation}
Additionally, we evaluate the inverse participation ratio (IPR) give by 
\begin{equation}\label{IPR}
    IPR = \left( \sum_j P_j^2 \right)^{-1}
     . 
\end{equation}
Both $S$ and  $IPR$ have two well-defined extremes.  For a wavepacket entirely distributed across the $N$ sites: $S = \log N$ and $IPR=N$. On the other hand, for a wavepacket fully localized: $S=0$ and $IPR=1$.

The results for the Shannon entropy are shown in Fig.~\ref{fig:Shannonvst}. We see that our Parrodian QW (line associated with $\omega_m$) has a wavepacket more distributed across the lattice than the cases with weak spreading 
(lines  associated with $\beta_1$ and $\beta_2$).  However, we observe that the defect-free QW produces more Shannon entropy than our Parrodian QW, indicating that it is more delocalized that our model.
This behavior is further confirmed with the results of the $IPR$, as presented in Fig.~\ref{fig:IPRvst}. 
The weak spreading (for $\beta_1$ and $\beta_2$)
is nearly localized, in contrast to the defect-free QW, which is significantly distributed across the lattice. Between both cases we observe our Parrondian QW with an intermediate $IPR$.
These results show that both $S$ and $IPR$ are in agreement 
and highlight that the Parrondian QW is accompanied by a wavepacket distributed over fewer sites across the lattice when compared to the defect-free QW.

To gain deeper insights into the underlying mechanisms of our Parrondian QW, we conducted a comprehensive analysis of its associated probability distribution. In the panels (a) and (c) of Fig.~\ref{fig:prqd} it is evident that the central region of the defect-free QW distribution has a higher probability of being populated than the corresponding region of the Parrondian QW distribution. Such result is in agreement with the insights obtained from the Shannon entropy and IPR.

The wavepacket spreading is evaluated by the standard deviation, which is a global measure (assessed for the entire lattice). Thus, let us analyze a local version of this measure.
In order to  grasp local contributions for the spreading we compute the Relative Quadratic Deviation (RQD)~\cite{pires2021negative}
\begin{equation}\label{RQD}
    RQD(j) = \left( j-\overline{j} \right)^2 P_j  .
\end{equation}
It is evident from 
panels (b) and (d) of Fig.~\ref{fig:prqd}  
that the local contributions to the global standard deviation 
are primarily determined by the peaks at the edges of the probability distribution. Comparing the maximum RQD of Parrondo case $RQD_{mp}$ with defect-free case $RQD_{md}$, we obtain $ RQD_{mp}/RQD_{md} \approx 4.44 $, which is in accordance with the respective ratio of maximum probabilities $ P_{mp} / P_{md} \approx 4.47 $. 

In summary, all the aforementioned insights collectively indicate that the enhancement of the spreading observed in the Parrondian QW stems from  a reduction in probability within the central region, resulting in a relative accumulation at the border of the QW distribution.

\section{Final remarks}\label{sec:remaks} 

We conducted an investigation into the transport properties of CTQWs in the presence of time-dependent transition defects. Our model was formulated to account for alternating configurations in which these defects traditionally play a role in reducing the wavepacket dispersion.

Our results reveal the first manifestation of a Parrondian effect in the domain of CTQWs. 
In our protocol, we show that the alternating use of two unfavorable setups, where defects decelerate wavepacket spreading, can lead to scenarios in which the wavepacket spreads faster than the defect-free CTQW.

Our findings offer a fresh perspective on how to use defects, which are usually seen as detrimental, to improve quantum transport. This novel approach has the potential to enhance the efficiency and reliability of quantum transport systems, making our results  promising for future developments in the field of quantum transport.
As comprehensively exemplified in Sec.\ref{sec:intro}, QWs are connected to a plethora of phenomena in several domains of research, thus our results
can provide insights about the role of alternating defects in applied fields~\cite{ambainis2003quantum,portugal2013quantum,venegas2008quantum,venegas2012quantum,kadian2021quantum} as well as fundamental areas~\cite{kitagawa2010exploring,wu2019topological}.

In future works, we plan to investigate the influence of temporal switching, incorporating both positive and negative correlations, on the manifestation of the PE in CTQWs. As shown recently~\cite{pires2021negative} negative correlated temporal disorder is able to produce non-trivial effects in QWs.
The investigation of the effects of chaotic switching~\cite{lai2021chaotic} in our protocol is also an important research endeavor.

\section*{Acknowledgments}
We acknowledge the FAPEMIG for financial support.

\bibliography{main.bib}

\end{document}